\documentstyle[preprint,aps,amsfonts]{revtex}

\begin{document}
\title{Constrained systems described by Nambu mechanics}
\draft
\preprint{UM-P-96/37; RCHEP-96/4}
\author{C. C. Lassig\footnote{ccl@physics.unimelb.edu.au}
and G. C. Joshi\footnote{joshi@physics.unimelb.edu.au}}
\address{Research Centre for High Energy Physics, \\
School of Physics, The University of Melbourne, \\
Parkville, Victoria 3052. Australia}
\date{\today}
\maketitle
\begin{abstract}
Using the framework of Nambu's generalised mechanics,
we obtain a new description of constrained Hamiltonian
dynamics, involving the introduction of another degree of
freedom in phase space, and the necessity of defining the
action integral on a world sheet. We also discuss the problem
of quantising Nambu mechanics.
\end{abstract}

\section{Introduction}
In 1973, Nambu introduced a dynamics similar to the classical Hamiltonian
formalism, but generalised to a phase space of three dimensions \cite{Nambu}.
Many authors have since investigated its application, and connection to
standard mechanics \cite{Misc}. Recently, Takhtajan \cite{Takh} has
generalised the concept to higher dimensional phase spaces, and obtained
algebraic criteria for this form of dynamics, as well as a Principle of 
Least Action to give the equations of motion. 

In this letter we aim to show how Hamiltonian dynamics with constraints
can be written in terms of Nambu mechanics. In this formalism, it is possible
to describe systems in which the constraints are incompatible with the
Hamiltonian.  In Section I we introduce the principles of Nambu mechanics,
and we  develop our application in Section II. In Section III we look
at the problem of quantising Nambu mechanics, and consider the properties
that the hypothetical algebra of operators must satisfy.

\section{Fundamentals of Nambu mechanics}
Nambu mechanics \cite{Nambu} on the three dimensional phase space 
${\Bbb R}^3 = \{ (x,y,z) \}$ is described by the pair of hamiltonian 
functions $H_1$ and $H_2$, with the motion of any function $F(x,y,z)$
given not by a Poisson bracket, but instead by the ternary {\em Nambu bracket},
\begin{equation}
\frac{dF}{dt} = (F,H_1,H_2) = \frac{\partial(F,H_1,H_2)}{\partial(x,y,z)}.
\label{Nambu}
\end{equation}
The Nambu bracket is a trilinear map with the following properties:

\begin{enumerate}
\item it is antisymmetric under exchange of any of its arguments
\begin{equation}
(A,B,C) = -(B,A,C) = -(A,C,B) = -(C,B,A),
\end{equation}

\item it is a derivation (Leibniz rule)
\begin{equation}
(A,B,CD) = C(A,B,D) + (A,B,C)D,
\end{equation}

\item it satisfies Takhtajan's Fundamental Identity \cite{Takh}
(analogous to the Jacobi identity for Poisson brackets)
\begin{equation}
((A,B,C),D,E) + (C,(A,B,D),E) + (C,D,(A,B,E)) = (A,B,(C,D,E)).
\end{equation}
\end{enumerate}

The velocity vector of the flow $g^t$ in phase space is
$^*(dH_1 \wedge dH_2) = \vec{\nabla} H_1 \times \vec{\nabla} H_2$, which
means that the Nambu bracket can be written as
\begin{equation}
(F,H_1,H_2)(x) = \frac{d}{dt} \Big|_{t=0} F(g^t x) 
= dF(^*(dH_1 \wedge dH_2)).
\label{flow}
\end{equation}
The velocity vector is chosen to give the relation
$(dH_1 \wedge dH_2)(\vec{\xi_1}, \vec{\xi_2}) =
\omega^3(\vec{\nabla} H_1 \times \vec{\nabla} H_2, \vec{\xi_1}, \vec{\xi_2})$,
so that $\omega^3 \equiv dx \wedge dy \wedge dz$
is an absolute integral invariant
of the phase flow, ensuring Liouville's theorem, the invariance of phase
space volume under canonical transformations.

Takhtajan \cite{Takh} has shown that the 2-form 
$\omega^2 = x dy \wedge dz - H_1 dH_2 \wedge dt$ plays the role of a
generalized Poincar\'e-Cartan integral invariant, from which can be 
constructed a Principle of Least Action, in which the action,
defined as the integral of $\omega^2$ over a 2-chain, i.e. the world sheet
of a closed string, is extremal for a flow $g^t$ satisfying the Nambu equations
of motion. If we use a parametrisation which introduces a string variable $s$,
$0 \leq s \leq 1$, and a notation in which a dot represents the time
derivative, and a prime represents differentiation with respect to $s$, then
the action can be written:
\begin{equation}
I = \int_{t_0}^{t_1} \int_0^1 \Big[ x (y^{\prime}\dot{z} - \dot{y} z^{\prime})
- H_1 \Big( \frac{\partial H_2}{\partial x} x^{\prime}
+ \frac{\partial H_2}{\partial y} y^{\prime}
+ \frac{\partial H_2}{\partial z} z^{\prime} \Big) \Big] ds \, dt.
\label{action}
\end{equation}

\section{Constrained dynamics in Nambu form}
To start with, we consider a two-dimensional phase space with Hamiltonian
$H(q,p)$, and one constraint, $\phi(q,p) = 0$. Then the equations of motion are
\begin{equation}
\frac{dq}{dt} = \frac{\partial H}{\partial p} 
+ \lambda \frac{\partial \phi}{\partial p}; \; \;
\frac{dp}{dt} = - \frac{\partial H}{\partial q}
- \lambda \frac{\partial \phi}{\partial q},
\label{Dirac}
\end{equation}
where $\lambda$ is the Lagrange multiplier. Solving these equations is a matter
of finding the value for $\lambda$ for which the constraint is independent of
time, i.e. $d\phi/dt = 0$. 

To write this in a Nambu form, we need to introduce another phase space
variable, $r$. The two hamiltonians are then
\begin{equation}
H_1(q,p,r) = H(q,p) - r; \; \; H_2(q,p,r) = r + \lambda \phi(q,p).
\end{equation}
If we define the Nambu bracket as
\begin{equation}
(F,H_1,H_2) = \frac{\partial(F,H_1,H_2)}{\partial(q,p,r)},
\end{equation}
we obtain the same equations of motion for $q$ and $p$, Eq.~(\ref{Dirac}),
together with
\begin{equation}
\frac{dr}{dt} = \lambda \Big( \frac{\partial H}{\partial q}
\frac{\partial \phi}{\partial p} - \frac{\partial H}{\partial p}
\frac{\partial \phi}{\partial q} \Big)
= -\lambda \frac{d\phi}{dt}.
\end{equation}
The variable $r$ is an extra degree of freedom which we introduced
artificially in order to construct our Nambu formalism, so ideally we would
like it to decouple from the dynamics, i.e. we want $dr/dt = 0$. This requires
either $\lambda = 0$, i.e. the constraint is removed from the system, which
will certainly give a plausible result but not the one we were looking for,
or instead $d\phi/dt = 0$, which is what we wanted.

We can extend this construction to a higher dimensional phase space,
$\{ (q_i,p_i)\}$ ($i,j = 1,\ldots,n$), 
with $m$ constraints $\phi_k(q_i,p_i) = 0$ ($k=1,\ldots,m$),
by Nambu mechanics with hamiltonian functions
\begin{equation}
H_1(q_i,p_i,r) = H(q_i,p_i) - r; \; \; 
H_2(q_i,p_i,r) = r + \sum^{m}_{k = 1} \lambda_k \phi_k(q_i,p_i).
\end{equation}
The Nambu bracket is given by
\begin{equation}
(F,H_1,H_2) = \sum^n_{i=1} \frac{\partial(F,H_1,H_2)}{\partial(q_i,p_i,r)}.
\end{equation}
Once again, we try to decouple the extra degree of freedom by solving for
$dr/dt = 0$, without putting the Lagrange multipliers equal to zero.

It is here that we find a use for this formalism. In most dynamical systems
considered in physics the constraints are consistent with the Hamiltonian,
and solving for the Lagrange multipliers is possible. This is not always the
case however, and standard Hamiltonian mechanics cannot cope with such systems.
In the Nambu formalism, though, we can see that this corresponds to a situation
in which it is not possible to decouple the extra degree of freedom.
Thus the Nambu formalism combines the influence of both the Hamiltonian
$H(q_i,p_i)$ and the constraints, at the price of introducing an extra variable
to phase space, and having to define the action on a world sheet instead of
a world line. 

An example of such a case is that of the octonionic field theory with
electromagnetic duality that we have considered in an earlier paper \cite{Mag}.
This theory allows for both electric and magnetic charges by using a
nonassociative octonionic field, but the nonassociativity prevents us
from solving the constraint equations.

\section{Algebraic requirements for quantisation}
In quantising standard Hamiltonian mechanics, the Poisson bracket between
observables is replaced by the algebraically equivalent Lie bracket or
commutator between operators. For Nambu mechanics then, we require an
algebra which possesses a ternary product satisfying the conditions 1--3
in Section I. Nambu has already shown that this cannot be done completely
with an associative algebra, so we look towards nonassociative structures
\cite{Schafer}.

Alternative algebras, such as the octonions, for which $x(xy) = x^2 y$ and
$(yx)x = y x^2$, possess derivations, necessary to satisfy condition 2,
of the form
\begin{equation}
D_{a,b} x =  (ax)b - a(xb) + b(ax) - a(bx) + (xa)b - (xb)a,
\end{equation}
where $a,b,x$ are all elements of the alternative algebra. This derivation
is symmetric in $a$ and $b$, but not for an exchange of either with the
argument $x$. The same applies for the derivations of the Jordan algebras,
which are commutative but not associative, and their elements satisfy the
property $(xy)x^2 = x(yx^2)$. The derivations of such algebras are of the form
\begin{equation}
D_{a,b} x = (ax)b - a(xb).
\end{equation}
There are few other nonassociative algebras that have been studied in much
detail, and none have yet been found to possess the necessary properties.

The difficulty can perhaps be seen if we consider that a Poisson bracket
$(A,B)$ is a linear transformation $L_B A$, the Lie derivative of $A$
in the direction of the phase velocity defined by the hamiltonian function $B$.
These Lie derivatives then, naturally enough, make up a Lie algebra.
The Nambu bracket $(A,B,C)$ can be considered as a bilinear transformation
on $(A,B)$, but it is not possible to make an algebra out of those 
transformations. Alternatively we could define the Nambu bracket as a 
Lie derivative of $A$ in the direction $\vec{\nabla} B \times \vec{\nabla} C$,
but then it is not possible to have separate elements of the algebra for each of
$B$ and $C$, which would be necessary if we were to treat them as separate
operators in a quantum theory.

It is still possible though that there is a nonassociative algebra satisfying
the necessary criteria, enabling a quantum Nambu mechanics on the same
footing as the quantum Hamiltonian formalism.

\end{document}